\documentclass[preprint]{aastex}
\usepackage{natbib}
\usepackage{paralist}
\usepackage{adjustbox}
\usepackage{graphicx}

\usepackage[normalem]{ulem}

\shorttitle{No Giant Planet Pileup Near 1 AU}
\shortauthors{Wise \& Dodson-Robinson}

\begin{document}


\title{No Giant Planet Pileup Near 1 AU}

\author{A. Wise and S. Dodson-Robinson}
\affil{University of Delaware}

%\author{A. W. Wise\altaffilmark{1,2} and S. E. Dodson-Robinson\altaffilmark{1,3}}
%\affil{University of Delaware, Newark, DE 19716}

%\altaffiltext{1}{University of Delaware, Department of Physics and Astronomy, 217 Sharp Lab, Newark, DE 19716, USA}
%\altaffiltext{2}{Email: aww@udel.edu}
%\altaffiltext{3}{Email: sdr@udel.edu}

\clearpage

%\begin{abstract}
%A pileup near 1~AU in the semimajor axis distribution of giant exoplanets
%has been visually identified using log-spaced distribution plots. Research
%efforts have attempted to find a physical explanation for this pileup. Here we
%propose that looking for features in a log-spaced semimajor axis
%distribution of giant planets is problematic. We use the Bayesian Blocks
%algorithm to analyze the linear-spaced semimajor axis distribution,
%and find that the apparent pileup is not significant.
%\end{abstract}


%\keywords{planets and satellites: gaseous planets, planets and satellites: fundamental parameters, planets and satellites: dynamical evolution and stability, planet–disk interactions}

\clearpage

\section{Introduction}

Using semimajor axis distributions with logarithmic spacing, many investigators \citep{us07,w09,hp12,bn13} have suggested that there exists a pileup (or other distribution feature near 1~AU, which we hereafter refer to as a pileup) in the distribution of giant exoplanets near 1-AU.
%A pileup has been identified near 1~AU in the log-spaced semimajor axis
%distribution of giant exoplanets from radial velocity (RV) surveys.
%It has been proposed that protoplanetary disk
%clearing by photoevaporation may create such a pileup \citep{ap12,er15}.
%Following recent work suggesting photoevaporation does not create such a
%pileup \citep{wd18},
We propose that the pileup (or any distribution feature) is not significant.
Since we do not have knowledge of individual RV surveys' many
complex sampling and selection biases,
we cannot assign a proper statistical significance to the pileup.
%However, as fully incorporating the different biases from each RV survey
%has not been accomplished (to our knowledge) and would require information
%about target selection and time sampling from each planet-search team,
Given access to the publicly available data only,
analyses akin to the following may be the best way to assess whether the
claimed pileup exists in the current set of confirmed exoplanets.

The mass-semimajor axis distribution of RV-detected giant planets\footnote{\label{note1}Planet orbit and mass data from exoplanet.eu/catalogs, March 16, 2017 \citep{s11}} is shown
on a logarithmic semimajor axis scale (top-left), and
a linear scale (top-right) truncated at 2.5 AU to avoid
observational biases against detecting long-period planets.
From these plots, it is clear that giant planet semimajor axes are not log-uniformly distributed, a fact that partly triggered the pileup proposal. However, we assert that the {\it expected} exoplanet semimajor axis distribution should {\it not} be log-uniform. A simple argument is as follows:
When looking for non-uniform features in the
distribution of log($a$), the null hypothesis is that the giant planet
frequency scales with $1/a$, as the plot interval allotted to each AU of
semimajor axis goes as $1/a$. According to \cite{a07}, \cite{bk11a,bk11b},
and \cite{cn14}, the majority of giant planets within 5~AU probably underwent
significant inward migration after forming.
If they migrated at local viscous timescale,
$r^2/\nu$, and we assume $\nu(r) \sim r$ \citep[e.g.][]{h98}, these planets would
migrate slower at larger $a$, so giant planet frequency per AU would
{\it increase} with $a$.
%The simplest model for such inward migration of giant
%planets is that the migration timescale of the planet is the local
%viscous timescale, $r^2/\nu$. Using the expected form of viscosity $\nu
%\sim r$, this simple scaling suggests giant planets will migrate slower
%at larger $a$ than at smaller $a$, and hence the giant planet frequency
%per AU will actually {\it increase} with $a$. While this viscous
%timescale model is far too simple to describe actual planet migration,
%it shows us how important the axis scale is when looking for visual
%features in a distribution.

\section{Bayesian Blocks Analysis}

Here we use the Bayesian Blocks algorithm \citep{s13} to assess the extent
to which the 1-D distribution of giant planet semimajor axes,
plotted on a linear scale, differs from a
uniform distribution. First we resample the exoplanet mass-semimajor axis
distribution using reported observational uncertainties\footnotemark[1]. For each planet and
each variable (mass, semimajor axis), we construct two Gaussian probability
density functions (PDFs), one from the upper and one from the lower
1-$\sigma$ error. We randomly select which PDF to use and
then pick a random value from its domain (with probability weighted by
PDF), adding this error value to the reported semimajor axis or mass.
For reported errors that were undefined or zero according to exoplanet.eu,
we assumed 1-$\sigma$ errors of 1\% of the planet's
reported mass or semimajor axis---much tighter than typical
reported error bar measurements \citep{s11}. We resampled the
mass-semimajor axis distribution 1000 times, then computed each
corresponding 1-D semimajor axis distribution for planets with resampled
$M \sin i > 0.5 M_{\rm Jup}$. The ensemble of 1000 1-D semimajor axis
distributions for giant planets forms the solid blue histogram shown in
the middle-left.

Now, we use the Bayesian Blocks algorithm to construct optimally binned semimajor axis histograms of giant exoplanets \citep{s13}:
We adopt a geometric prior on
the number of blocks (histogram bins), $P(N_{\rm blocks}) = P_0
\gamma^{N_{\rm blocks}}$, and choose a threshold probability for correct
detection of all bin edges (``change points'') of $p_* = 0.95$. For each
resampled semimajor axis distribution A containing N$_{\rm A}$ planets,
we compute an appropriate $\gamma$ using a set of random datasets
R$_{\rm N_A}$ (all with N$_{\rm A}$ planets) drawn from a uniform semimajor
axis distribution. First, for each of a large number of R$_{\rm N_A}$, we find the
largest $\gamma$ for which the Bayesian Blocks algorithm applied to
R$_{\rm N_A}$ finds a number of change points less than or equal to $(1
- p_*) ({\rm N_ A} - 1)$, and average over these values of
$\gamma$. Second, we use this average $\gamma$ in our prior as we use
the Bayesian Blocks algorithm to find the number of change points in A,
$N_{cp}$. These two steps are repeated, only changing the maximum number
of change points detected to $(1 - p_*^{1/N_{cp}}) ({\rm N_ A} -
1)$ using the value of $N_{cp}$ from each previous iteration, until
$N_{cp}$ stops changing. This iterative algorithm does not converge if a
dataset A is truly uniform, with $N_{cp}=0$, and we note 83 out of
1000 resampled exoplanet semimajor axis distributions did not converge
within $>100$ iterations, so we treat these 83 distributions as having
zero change points. The two bottom plots show sample semimajor axis
histograms produced by the Bayesian Blocks algorithm.

The middle-right shows a histogram of the histogram change-point
locations from our Bayesian Blocks analysis. The most notable feature is
78\% of resampled realizations of planet semimajor axes show higher
planet occurrence rate between 1 and 1.5 AU, indicated by a pair of change
points---one to increase the planet frequency (positive in plot),
and a second to decrease it (negative in plot).
As this feature only appears in 78\% of resampled semimajor axis distributions,
we suggest that the evidence for a pileup near 1~AU is not significant.

\begin{figure}
\centering
%\plottwo{f1a.eps}{f1b.eps}
%\plottwo{f2a.eps}{f2b.eps}
%\plottwo{f3a.eps}{f3b.eps}
%\plottwo{f1a-eps-converted-to.pdf}{f1b-eps-converted-to.pdf}
%\plottwo{f2a-eps-converted-to.pdf}{f2b-eps-converted-to.pdf}
%\plottwo{f3a-eps-converted-to.pdf}{f3b-eps-converted-to.pdf}
\plotone{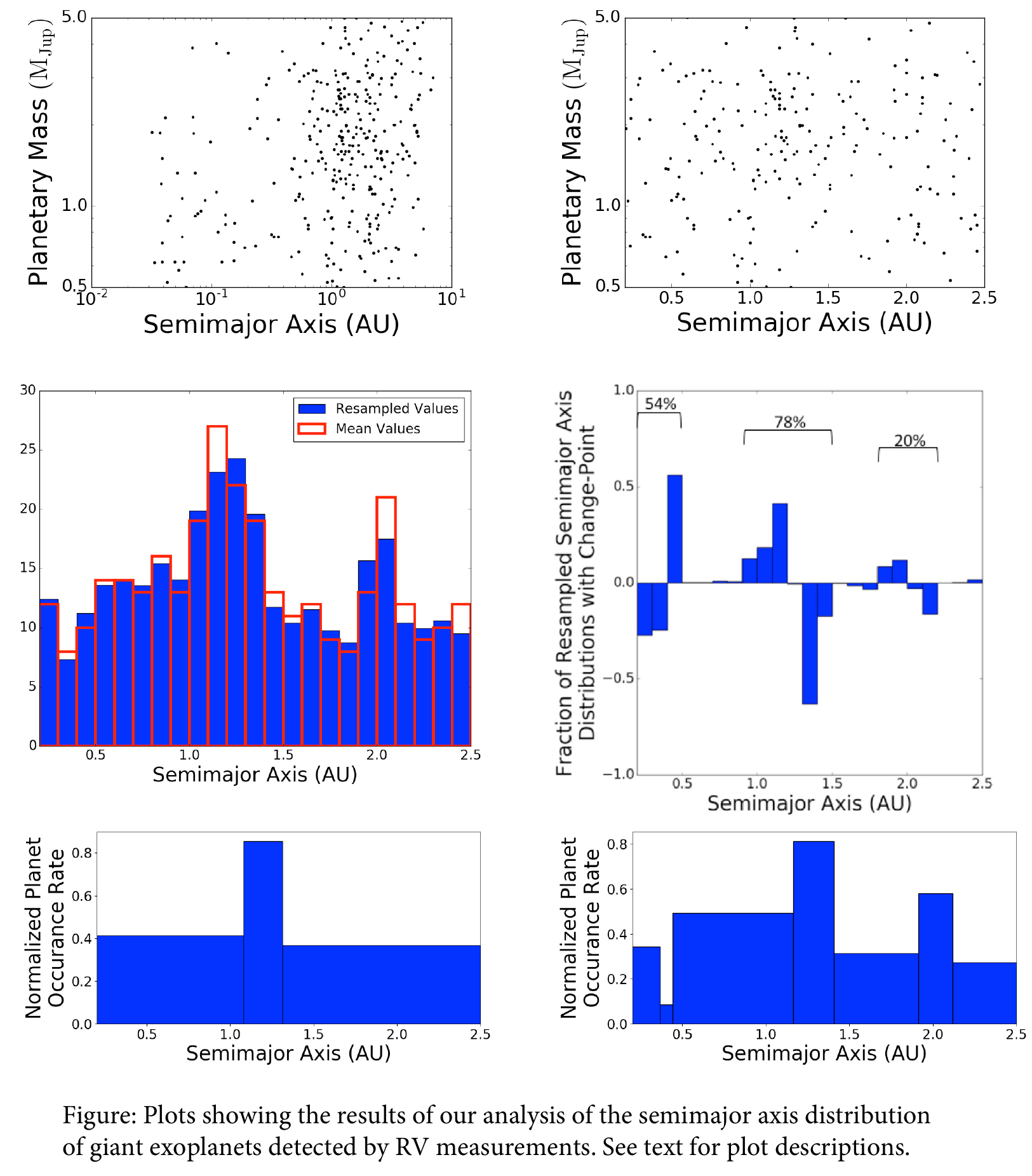}
%\includegraphics[width=0.7\linewidth]{}
%\caption{For plot descriptions, see text.\label{fig1}}
\vspace{1.0cm}
\end{figure}

\acknowledgments
The authors thank Eric Ford for useful ideas and acknowledge support from the UNIDEL foundation and NSF CAREER award 1520101.

\clearpage

\bibliography{pileup}
\clearpage

%% The following command ends your manuscript. LaTeX will ignore any text

\end{document}